# Engineering High-Order Harmonic Generation through Gas Confinement at Sub-Millimeter Lengths

Agata Azzolin[1,2,3], Gaia Giovannetti[1,2], Oliviero Cannelli[1,3], Sabine Rockenstein[1,2,4], Guangyu Fan[1,3,5], Md S. Ahsan[1,6], Lorenzo Colaizzi[1,2,7], Erik P. Månsson[1], Noah Tettenborn[1,2], Linda Oberti[1,6,7], Davide Faccialà[6], Fabio Frassetto[8], Anna Gabriella Ciriolo[6,7], Dario W. Lodi[7], Alia Ashraf[6,7], Cristian Manzoni[6], Rebeca Martínez Vázquez[6], Michele Devetta[6], Roberto Osellame[6,7], Luca Poletto[8], Salvatore Stagira[6,7], Caterina Vozzi[6], Terry Mullins[1,3], Vincent Wanie[1], Andrea Trabattoni[1,9,10], Francesca Calegari[1,2,3]*.

1. Centre for Free-electron Laser Science, Deutsches Elektronen-Synchrotron, Notkestr. 85, 22607 Hamburg, Germany
2. Physics Department, University of Hamburg, Luruper Chausee 149, 22761 Hamburg, Germany
3. The Hamburg Centre for Ultrafast Imaging, University of Hamburg, Luruper Chausee 149, 22761 Hamburg, Germany
4. Max Planck Institute for the Structure and Dynamics of Matter, Luruper Chaussee 149, 22761 Hamburg, Germany
5. Shangai Key Lab of Modern Optical System, University of Shangai for Science and Technology, Shangai 2000093, China
6. Institute for Photonics and Nanotechnologies, Consiglio Nazionale delle Ricerche, piazza L. da Vinci 32, 20133 Milano, Italy
7. Physics Department, Politecnico di Milano, piazza L. da Vinci 32, 20133 Milano, Italy
8. Institute for Photonics and Nanotechnologies, Consiglio Nazionale delle Ricerche, via Trasea 7, 35131 Padova, Italy
9. Institute of Quantum Optics, Leibniz Universität Hannover, Welfengarten 1, 30167 Hannover, Germany
10. Cluster of Excellence PhoenixD (Photonics, Optics, and Engineering-Innovation Across Disciplines), Leibniz Universität Hannover, Welfengarten 1, 30167 Hannover, Germany

*Corresponding author: francesca.calegari@desy.de

**Abstract**

Attosecond light sources based on high-order harmonic generation (HHG) constitute to date the only table-top solution for producing coherent broadband radiation covering the spectral range from the extreme ultraviolet to the soft X-rays. The so-called emission cutoff can be extended towards higher photon energies by increasing the driving wavelength at the expense of conversion efficiency. An alternative route is to overdrive the process by using higher laser intensities, with the challenging requirement of interacting with higher plasma densities over short propagation distances. Here, we address this challenge by using a differentially pumped glass chip designed for optimal gas confinement over sub-mm lengths. By driving HHG with multicycle pulses at either 800 nm or 1500 nm, we demonstrate a cutoff extension by a factor of two compared to conventional phase matching approaches and surpassing the present record using multicycle fields. Our three-dimensional propagation simulations, in excellent agreement with the experiment, confirm that gas confinement is crucial since efficient phase matching of cutoff harmonics occurs only for short propagation lengths. Additionally, we show that the high photon energy component is not only temporally confined to the leading edge of the driving pulse, but also spatially confined in the near-field to an off-axis contribution due to reshaping of the driving field along propagation inside the medium. Our findings contribute to the fundamental understanding of HHG across different regimes.

**Introduction**

The ever-growing applications of HHG table-top attosecond sources in microscopy and imaging [1–4], attosecond spectroscopy [5–11], and quantum optics [12,13] call for their optimization in terms of spectral extension (the so-called cutoff emission) and brightness. This is particularly challenging when targeting the generation of soft X-rays by driving HHG with the direct output of a near-infrared laser. On the one hand, the cutoff energy scales linearly with the intensity, $I$, of the driving field [14]. On the other hand, the plasma density resulting from the ionization of the generation medium at high intensities affects the spatial and temporal properties of the driving field along the propagation, limiting the effective phase matching length.

In the past years, several strategies have been proposed to achieve perfect phase matching at high photon energies. Conventional generation schemes rely on waveguides or loose focusing geometries [15–23], for which the intensity of the driving field in the generation medium is tuned below the critical ionization threshold. In this geometry, the laser intensity is either constant or slowly varying along the propagation direction (gas medium length << Rayleigh range), leading to the so-called adiabatic regime. Other approaches rely on quasi-phase matching (QPM), where the phase mismatch is compensated along the propagation direction by either modulating the driving laser phase and intensity, or the medium density [24–28]. To reach photon energies above 100 eV, these schemes additionally rely on mid-infrared driving wavelengths, since the cutoff scales quadratically with the driving wavelength $\lambda$ [14]. However, this strategy often requires optical parametric amplification (OPA) or optical parametric chirped pulse amplification (OPCPA) sources, which substantially increase the complexity of the driving laser system. Moreover, longer driving wavelengths drastically reduce the HHG yield, as the conversion efficiency scales as $\lambda^{-5.5}$ to $\lambda^{-6.5}$ [29,30].

Circumventing these limitations requires substantial changes in the generation scheme, as in the case of the so-called overdriven or nonadiabatic regime [31–42]. In this regime, intensities in the order of $10^{15}$–$10^{16}$ W/cm$^2$ and a tight focusing geometry (gas medium length greater or comparable to the Rayleigh range) are employed to balance the phase matching terms (for a detailed discussion of the phase matching terms see Supplement 1, Section 2). In contrast to the conventional regime, the tight focusing geometry introduces a strong dipole phase matching term that scales with the intensity gradient, partially compensating for the dominant plasma contribution caused by the high driving field intensities. Additionally, propagation in a highly ionized medium leads to a strong temporal, spectral, and spatial reshaping of the driving field. Typically, reshaping effects are considered as detrimental, while here they transiently modify the generation conditions such that extended cutoff energies can be reached. These conditions, however, constrain the effective phase matching for high photon energies over short propagation distances (< 1 mm), making gas confinement the limiting factor for cutoff extension.

In this work, we overcome this challenge by using a newly designed differentially-pumped glass chip [43]. This cell provides highly efficient gas confinement, naturally reducing the interaction with the gas to sub-mm lengths and improving the HHG yield at the cutoff energies. We demonstrate extended cutoff energies matching, or even surpassing, the expected results [33–36] by using multi-cycle (30 fs) driving pulses, in contrast with previous works often stressing the importance of a few-cycle driver [31–33,36,39]. Optimization of the gas pressure and gas target position with respect to the driving beam focus is key to properly balancing plasma, spatial, and dipole phase contributions. We explored the generation in argon, neon and helium, both with 800 nm and 1500 nm driving wavelengths, showing remarkable cutoff extensions in all cases. Moreover, we employed a three-dimensional (3D) propagation model [44], based on the strong-field approximation [45], to prove the importance of short interaction lengths. Our results provide novel insights into the overdriven regime, showcasing the use of an engineered gas target for the generation of temporally and spatially confined radiation at high photon energies.

**Results**

A schematic layout of the gas target used for this experiment is shown in Fig. 1. The glass chip used to confine the gas is a modified version of the one used in [43] for third harmonic generation, and it is composed of a central gas reservoir of 960-μm length followed by a two-stage differential pumping sections. It was fabricated using the femtosecond laser irradiation followed by chemical etching (FLICE) technique [46,47]. A detailed description of the chip is reported in Supplement 1, Section 1. The gas distribution along the chip has been simulated using the free-molecular flow model in COMSOL Multiphysics [48]. When an input pressure of 250 mbar of neon gas is injected into the cell, as shown in the inset of Fig. 1, a pressure drop of approximately four orders of magnitude is predicted in the first pumping stage, *i.e.*, 1 cm away from the center, which is consistent with the experimentally measured pressure. The cell is placed on a motorized stage controlling three translation axes and one in-plane rotation axis, enabling the optimization of the generation process through precise control of the cell position with respect to the laser focus position. In our scheme, optimal conditions are achieved with the medium placed approximately one Rayleigh length after the focus and at gas pressures of a few hundred millibars. Positioning the medium further downstream reduces the intensity below the threshold for effective overdriven generation, while higher pressures and shorter distances cause excessive ionization early in the medium, preventing phase matching and suppressing the harmonic yield at high photon energies.

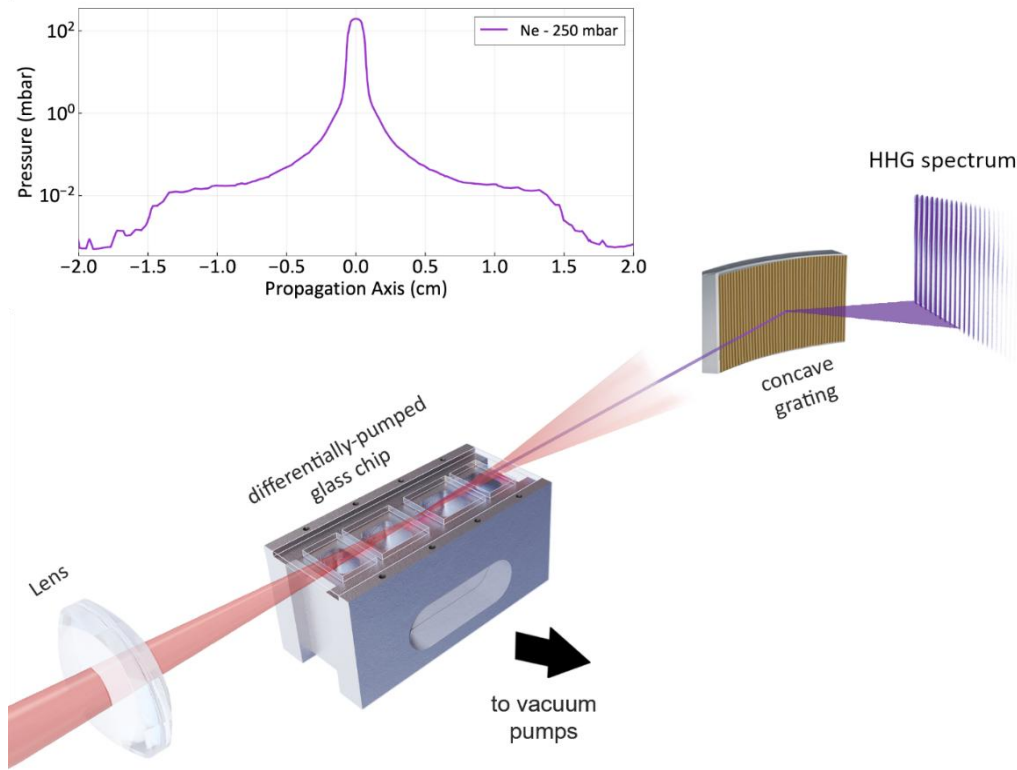

Fig. 1. Schematic layout of the HHG setup. The near infrared light (800 nm or 1500 nm) is tightly focused into the differentially pumped glass chip granting intensities in the order of $10^{16}$ W/cm$^2$. The harmonic radiation is then dispersed through a concave grating and collected at the detector. Inset: pressure profile along the cell (4 cm length) for the case of 250 mbar of neon input pressure as calculated using the free-molecular flow regime in the COMSOL simulation tool. The first pumping stage provides a pressure drop of four orders of magnitude when pumped at a speed of ca. 40 m$^3$/h corresponding to our experimental conditions.

Harmonic spectra generated in argon (top), neon (middle) and helium (bottom) with 800 nm (panels a-c) and 1500 nm (panels d-f) driving fields are reported in Fig. 2a-f. The peak intensity at focus for both driving wavelengths is on the order of $10^{16}$ W/cm$^2$, corresponding to a spot-size (1/e$^2$ radius) of 20 μm and pulse durations of 30 fs (see Supplement 1). The optimal gas pressures are reported in the figure for each case and range from 200 to 1800 mbar. Both driving wavelengths generate harmonic spectra extending far beyond conventional cutoff photon energies, which are indicated by the shaded areas in Fig. 2a-f and calculated from [37]. In all generating media, the measured cutoff energy (photon energy at which the HHG yield decreases to 1% with respect to its maximum) is approximately twice what is typically reported in the conventional regime. We note that, in contrast to standard generation approaches using Ti:Sapphire lasers, a photon energy of 92 eV can be accessed in the plateau region of the HHG spectrum generated in neon (Fig. 2b). This corresponds to a wavelength of 13.5 nm, which is particularly relevant for extreme ultraviolet (XUV) lithography [49,50].

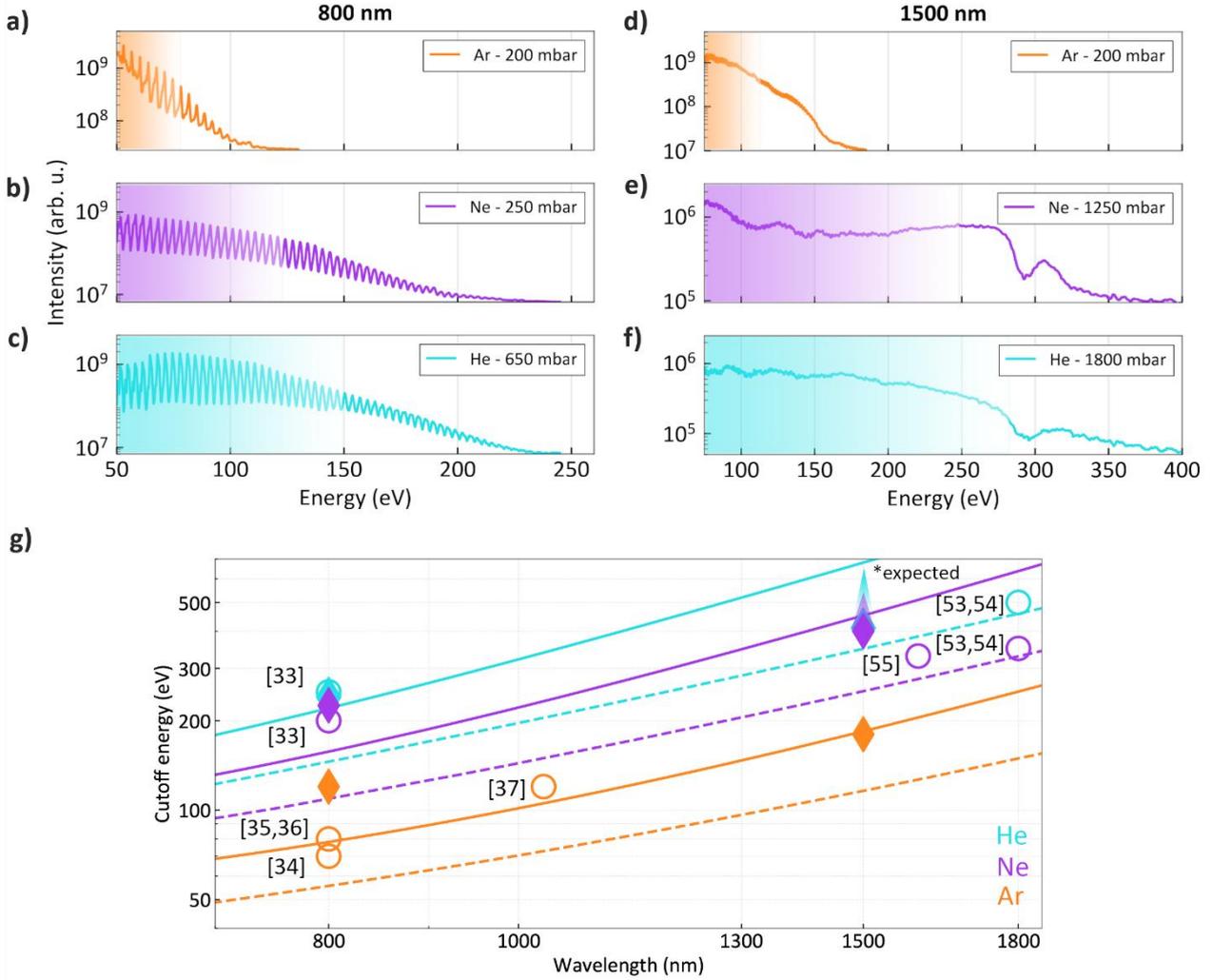

Fig. 2. (a-f) HHG spectra obtained in Ar (orange), Ne (purple) and He (light blue) with 800 nm (left) and 1500 nm (right) driving fields. The shaded areas in (a-f) highlight the expected cutoff in the conventional regime. (g) Theoretical curves (log-log scale) of the cutoff energies as a function of the driver laser wavelength in Ar (orange), Ne (purple) and He (light blue) in the conventional (dashed lines) and in the overdriven regime (full lines) from [37]. The circles mark the cutoff energies reported in literature, while the diamonds represent the results obtained in the current work. The markers for He and Ne with the 1500 nm driver underestimate the actual cutoff (see main text) and are reported as shaded areas towards higher energies to indicate qualitatively expected values.

In contrast to the spectra with the 800 nm driving field, the HHG spectra obtained with the 1500 nm exhibit a quasi-continuum (Fig. 2d-f). The effect cannot be attributed to the spectrometer resolution, which is better than 200 meV, but it can be explained by the physics of the HHG process. In fact, under optimal conditions, HHG driven by longer wavelengths is characterized by longer electron excursion times inducing stronger dipole phase contributions, which together with a higher degree of ionization, result in a stronger sub-cycle temporal reshaping of the driving field compared to the generation with shorter wavelengths [18,51]. These effects lead to spectral smearing of the discrete structure and the appearance of a quasi-continuum spectrum. As in the previous case, the cutoff energy in argon extends by almost a factor of two compared to the conventional regime. The spectra in neon and helium reach even higher energies, showing an absorption feature around 290 eV corresponding to the C K-edge due to carbon contamination of the beamline optics. The current spectrometer allows the detection of up to 400 eV, even though the grating is optimized only for energies up to 250 eV [52]. If higher photon energies could be recorded, the cutoff would likely extend even further, as a decrease of only an order of magnitude relative to the intensity maximum is observed in Fig. 2e-f. These results indicate the possibility of entering the water window without the need of using driving fields of 1800 nm or even longer wavelengths, so far privileged in the literature [53–56].

An overview of the cutoff energies reached as a function of the driving wavelength in argon, neon and helium is reported in Fig. 2g. The dashed lines correspond to the expected cutoffs in the conventional regime, while the solids lines show the cutoff in the overdriven regime; both curves are calculated from [37]. Our results are reported as diamonds, while the circles correspond to selected literature data [33–37,53–55]. In all cases, our generation conditions largely exceed the cutoff energies of the conventional regime. Furthermore, they either match or even surpass the expected values for the overdriven regime at 800 nm by 25 to 50%, reaching 120 eV, 225 eV and 240 eV for Ar, Ne, He, respectively. In particular, the 800 nm spectral cutoff in argon is the highest ever reported in literature, while in neon and helium we obtain extended cutoffs similar to Ref. [33], where the authors employed few-cycle pulses. In this respect, our work shows the notable advantage of using the direct output of the Ti:Sapphire laser (30 fs) to exceed 200 eV photon energies, simplifying the generation scheme for these photon energies. This demonstrates that the overdriven regime does not necessarily impose constraints on the pulse duration of the driving field, contrary to what was initially reported as a strict requirement from theory [31,32].

The HHG source driven with 800-nm pulses in argon and neon was also characterized in terms of flux. We observed experimentally that the condition yielding the maximum cutoff extension does not coincide with the condition of maximum brilliance. For argon, we measured 6.5 pJ total pulse energy, corresponding to about $6.8 \cdot 10^8$ photons/s at a central energy of 60 eV (within the 50-70 eV range). For neon, we measured a total pulse energy of 0.3 pJ, corresponding on average to $1.1 \cdot 10^7$ photons/s at a central energy of 150 eV (within the 50-200 eV range). These values underestimate the total flux of the source, as the toroidal mirror did not accommodate the beam's full vertical extent (see Supplement 1). We therefore expect that with further optimizations the flux could closely match values previously reported in the literature [32,39]. The spectra and photon fluxes presented here were obtained by solely maximizing the cutoff energy; however, further optimization of the gas pressure and medium position could enable a trade-off between cutoff extension and brilliance. This aspect will be the focus of future investigations.

To substantiate the claim that the exceptional gas confinement offered by the two-stage differentially pumped gas target is key to achieving the performances presented above, we performed simulations of the generation process based on a 3D full-space propagation model [44]. The model relies on the nonadiabatic strong-field approximation (SFA), which accounts for the response of each atom of the nonlinear medium to the full driving field, and thus to sub-cycle variations of the electron density. Additionally, the full-space simulation describes the sub-cycle variations of the driving field during the propagation in the medium, capturing the substantial reshaping of the NIR pulses due to propagation in the highly ionized medium characteristic of the overdriven regime. As proposed in Refs. [57,58], the ionization rate is computed using a modified version of the Ammosov-Delone-Krainov (ADK) model [59] (see Supplement 1). This model includes an empirical correction of the standard ADK model to take into account the higher ionization rates. This allows for matching the experimental results without the need to include contributions from multiple ionizations [39,60–62], and without implementing the computationally demanding above-barrier ionization (ABI) model [63], which would prevent performing full-space simulations. The calculations also include the absorption of the harmonic field through the gas medium, which was set to a constant gas density over the whole propagation length. We performed simulations for different media and driving wavelengths exploring a wide range of parameters, thus confirming the robustness of the model. In the following we will report on the results obtained in neon with the 800-nm driving wavelength by using the same parameters as in the experiment.

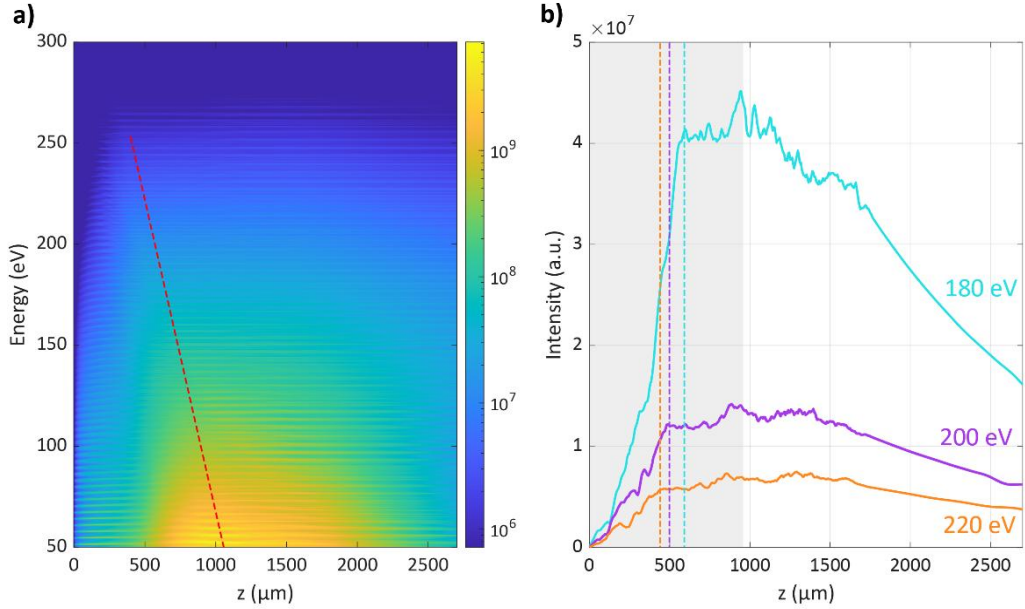

Fig. 3. Simulated high harmonic generation and three-dimensional space propagation over 2700 μm of the HHG spectrum in Ne with 800 nm driving pulse (a) Spectral intensity (atomic units, a.u.) integrated over the radial coordinate as function of the propagation axis (z) in the gas medium. The red dashed line highlights how saturation of the yield is progressively reached at shorter propagation distances, the higher the photon energy. (b) Intensity for three energies of interest (180 eV, 200 eV, 220 eV) as function of the propagation distance in the gas medium. The gray area corresponds to the actual medium length used in the experiment (900 μm). Intensity saturation (dashed vertical lines) is progressively reached at shorter propagation distances for higher HHG photon energies. Within the medium length, the yield of the cutoff energies is preserved.

**Discussion**

The relevance of gas confinement for efficient cutoff extension in the overdriven regime is highlighted in Fig. 3a, which shows the calculated harmonic spectrum generated by the 800-nm driver in neon and integrated over the radial coordinate as a function of an extended propagation distance of 2700 μm. While for photon energies below 100 eV, the harmonics yield progressively increases up to almost 1.5 mm, for higher photon energies saturation is reached well below 1 mm, thus within the gas target thickness of 900 μm used in the experiment and significantly decreases for larger propagation distances. For a more quantitative visualization of this effect, in Fig. 3b we report line cuts of the harmonic intensity for the selected values of 180 eV (light blue), 200 eV (purple) and 220 eV (orange). The gray area highlights the medium length used in the experiment. We identify three distinct regions along the propagation axis: a first region, extending up to approximately 500 μm, where the harmonic yield increases; a second region, up to about 1.5 mm, where the yield remains nearly constant; and a third, final region where the yield begins to decrease. The first region corresponds to a regime of effective phase matching, where the interplay between plasma, dipole, and geometric phase contributions favors the generation of harmonics at higher cutoff energies. Intensity saturation is reached progressively earlier in the medium for increasing photon energies, as indicated by the vertical dashed lines. This behavior is consistent with the decrease in coherence length at increasing harmonic order [64]. In the central part of the medium, phase matching conditions start to become unfavorable, yet reabsorption remains minimal [31], allowing the yield to be preserved. The length of the intermediate region decreases as absorption becomes dominant, a condition that is generally true the lower the photon energy (up to 100 eV), and therefore especially in the plateau region (see Supplement 1, Section 3). The step-like profile of the harmonic yield over sub-mm lengths is consistent with the description reported in Ref.[31] as signature of the overdriven regime. In the last region, harmonic propagation is increasingly affected by reabsorption, resulting in a yield reduction of nearly a factor of two over approximately 1 mm of propagation. This analysis reinforces the conclusion that efficient generation of extended cutoff emission in the overdriven regime strictly requires gas confinement on the sub-millimeter scale.

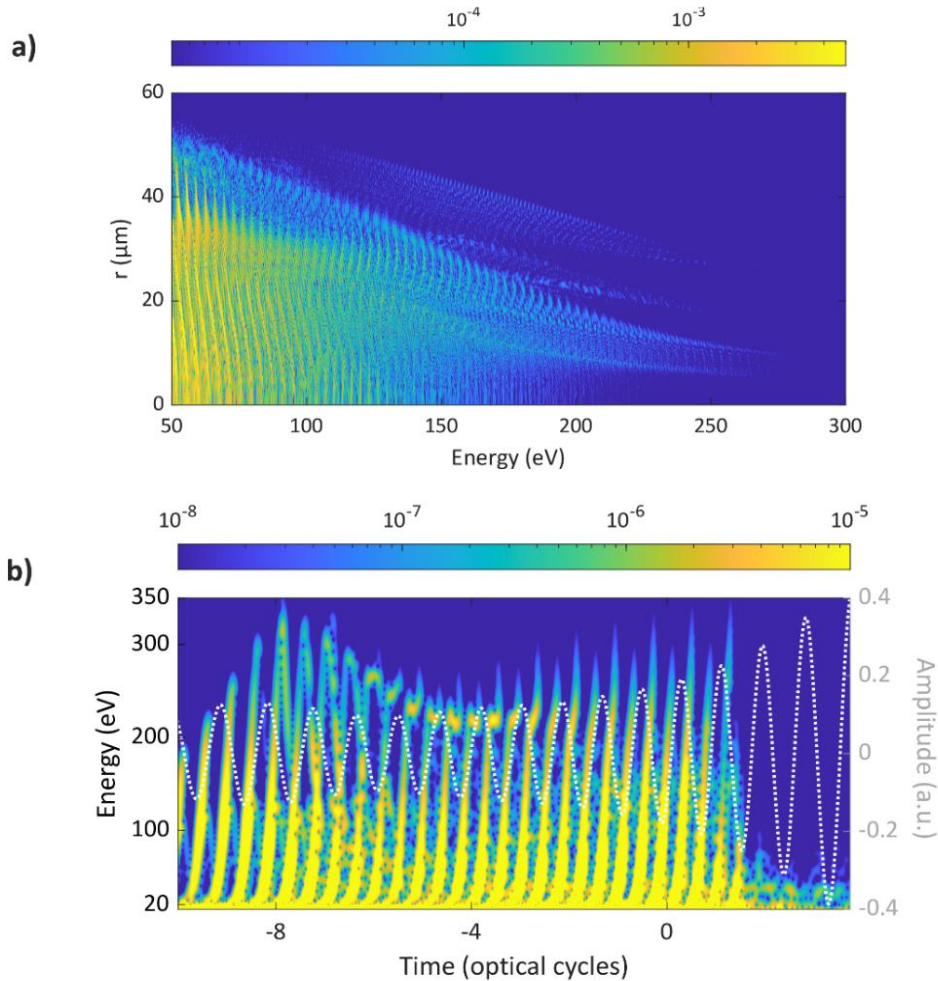

Fig. 4. Radial and temporal structure in the simulated harmonic spectrum at the end of the medium (z=900 µm), using Ne and a 800 nm driving pulse (a) HHG spectral intensity as function of the radial coordinate and photon energy. The high-energy cutoff is confined to radial coordinates between 5 and 15 µm, where 0 µm represents the on-axis propagation. (b) Gabor analysis for the spectrum in (a) for a radial coordinate of 8 µm, corresponding to conditions of maximum extension of the photon energy cutoff. The corresponding driving field amplitude is overlaid (white dashed line) to allow sub-cycle and envelope effects to be examined. The leading edge of the pulse is strongly reshaped, while harmonics are emitted preferentially with short trajectories from cycles -5 to 0.

Additional physical insights about the overdriven regime can be derived from a detailed analysis of the temporal and spatial characteristics of the emitted radiation. Fig. 4a shows the simulated spatial profile of the HHG spectrum as a function of the radial coordinate and photon energy computed at the end of the propagation in the medium (900 µm). Harmonics above 200 eV are generated off-axis and predominantly between 5 and 15 µm. This effect is a consequence of the full ionization of the medium in the central region of the beam (0-5 µm), which prevents efficient generation to take place (see Supplement 1). Notably, in the far-field this off-axis confinement is not observed, nor is it expected with our experimental geometry. We could in fact measure a rather homogeneous distribution of the harmonic spectra even in the cutoff region (see Supplement 1). Based on these results, we performed a Gabor analysis [65] of the harmonic spectrum at 8 µm radius, where the generation occurs for most of the cycles of the leading edge of the driving field (Fig. 4b). We find that the interplay between long and short trajectories varies at the sub-cycle time scale following the evolution and temporal reshaping of the driving field. For optical cycles between -10 and -8, corresponding to the maximum cutoff extension up to 300 eV, both short and long trajectories are equally contributing. For later cycles, the generation up to 220 eV is instead dominated by short trajectories. However, the region between -6 and -3 optical cycles presents interference trajectories to reconduct to a plasma-induced phase shift of the driving field (see Supplement 1, section 3). This peculiar generation dynamics can be explained by the transient nature of phase matching at high intensities, where both the plasma contribution and the intensity-dependent dipole phase vary on a sub-cycle timescale [66]. In neon, the higher ionization potential (21.6 eV) results in lower ionization levels. This reduces plasma dispersion and shifts the transient phase-matching conditions in favor of short trajectories, which exhibit weaker dipole phase sensitivity and are more robust against phase mismatch at higher harmonic orders. On the other hand, for the same generation conditions, in argon, characterized by a lower ionization potential (15.7 eV), long trajectories are favored at later cycles (see Supplement 1).

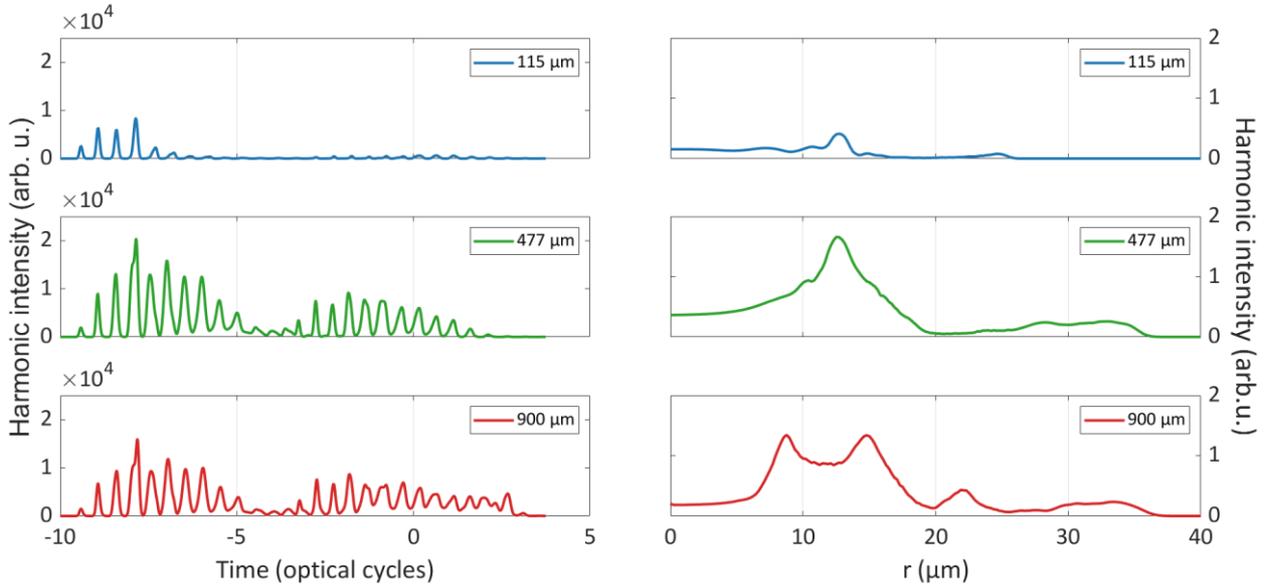

Fig. 5. Simulated harmonic intensity integrated between 200 and 220 eV as function of time (first column) and radial coordinate (second column), as obtained through Gabor analysis, at z = 115 µm (top), z = 477 µm (middle), z = 900 µm (bottom). The harmonic emission shifts towards longer time cycles, the longer the propagation distance. It is radially confined between 5 and 20 µm.

Moreover, we performed a propagation-dependent analysis showing a dynamic build-up of the harmonic field from different temporal and spatial regions across the medium. Fig. 5 reports the simulated harmonic intensity as per Gabor analysis integrated between 200 and 220 eV and over the radial coordinate as a function of time (first column), and over time as a function of the radial coordinate (second column). At small propagation distances (z = 115 µm), the high peak intensity of the driving field causes rapid plasma defocusing, limiting the phase matching to the first few cycles, before full ionization of the medium occurs (top panels). Additional propagation (z = 477 µm, middle panels) leads to off-axis maximization of the HHG intensity, with efficient generation shifting towards later time cycles. The strong plasma-induced reshaping of the driving field (see Supplement 1) leads to a shrinking of the leading-edge intensity, allowing for efficient phase-matching over multiple cycles. Finally, in the second half of the cell (z > 500 µm, bottom panels), the reshaping of the driving field turns into unfavorable phase-matching generation conditions, leading to an overall yield decrease in the central region of the beam and a spreading towards external radii.

Overall, our analysis clarifies the exceptional efficiency of our generation scheme, where we fully harness the spatiotemporal reshaping of the driving field in the overdriven regime by using a multicycle driver. The dynamic spatiotemporal confinement of the cutoff emission highlights the opportunity of further improving the generation conditions by engineering the spatial profile of the driving field to compensate for plasma defocusing and blue shifting of the central frequency, extending the overdriven regime beyond previous expectations.

**Conclusions**

In this work we present a newly designed glass chip characterized by highly efficiency gas confinement showing unprecedented cutoff energy extension in the overdriven regime using multicycle near-infrared driving pulses. By performing a 3D simulation of the HHG process that takes into account the experimental conditions, we show a dynamic reshaping of the driving field leading to a temporal and spatial confinement of the cutoff harmonics. Furthermore, our simulations rationalize the origin of increased phase matching efficiency for high-order harmonics using short generation media, highlighting the technological relevance of effective gas confinement over sub-mm scales and explaining the superior performances of our glass chip. By using the direct output of a commercial laser, our implementation strongly reduces the experimental complexities required for extended cutoffs, opening significant perspectives for simpler HHG schemes in both research and industry in the soft X-ray domain.

**Acknowledgements**

F.C. acknowledges funding from Cluster of Excellence 'CUI: Advanced Imaging of Matter' of the Deutsche Forschungsgemeinschaft (DFG)—EXC 2056—project ID 390715994, the Helmholtz-Lund International Graduate School (HELIOS) project number HIRS-0018, the Centre for Molecular Water Science (CMWS). O.C. acknowledges the Swiss National Science Foundation Postdoc.mobility program under the grant agreement P500PN_214151 and the European Union's Horizon Europe research and innovation program under the Marie Skłodowska-Curie METRICS HORIZON-MSCA-2022-PF-EF grant agreement no. 101106352. R.M.V. acknowledges the European Union's Horizon Europe research and innovation program under the FETOPEN grant agreement No 964588 (X-PIC). This research was supported in part through the Maxwell computational resources operated at Deutsches Elektronen-Synchrotron DESY, Hamburg, Germany. A.A. thanks Aksana Maria Wietrow for the support on the 3D graphic content.

**Disclosures**

The authors declare no conflict of interest.

**Data availability**

Data underlying the results presented in this paper are not publicly available at this time but may be obtained from the authors upon reasonable request.

**Supplemental document**

See Supplement 1 for supporting content.

# SUPPLEMENTARY INFORMATION

## S1 - Methods

The laser system used in this work is a carrier-envelope phase (CEP) stable Ti:Sapphire laser at 1 kHz repetition rate, with 800 nm central wavelength and 30 fs pulse duration (Femtopower). HHG was driven either using the direct output of the Ti:Sapph system (up to 1.5 mJ at the generation point), or the output of a custom-built two-stage CEP-stable optical parametric amplifier (OPA) [1] operating at 1500 nm, 30 fs duration, providing up to 500 µJ at the generation point. The driving fields were tightly focused into the gas target using a 15 cm focal length lens, yielding a beam waist of approximately 40 µm ($1/e^2$) in diameter, corresponding to peak intensities of $0.8–1 \times 10^{16}$ W/cm² and $4–5 \times 10^{15}$ W/cm², respectively.

The gas target is confined to a sub-mm length by employing a custom-designed differentially pumped chip. The laser-gas interaction region consists of a single channel of 300 µm inner diameter and 960 µm length, which is oriented along the laser propagation axis and placed on top of a vertical cylinder of 820 µm in diameter that is used as a gas reservoir. Moving away from the center, the channel scales to a diameter of 1.3 mm into two consecutive chambers of the chip having areas of 9x10 mm² and 9x6 mm² (1 mm separation distance). These chambers are directly connected to two separate vacuum pumps (nominal pumping speed 55 m³/h) ensuring differential pumping and a sudden pressure drop along the laser propagation axis. The cell is placed on a motorized stage controlling three translation axes and one in-plane rotation axis. A metering valve mounted in the gas line before the entrance to the chip allows fine tuning of the input gas pressure at the mbar level. A schematic of the chip dimensions is shown in Fig. S1 as per COMSOL Multiphysics simulations [2]. The total length of the chip is 41 mm, its thickness is 3 mm.

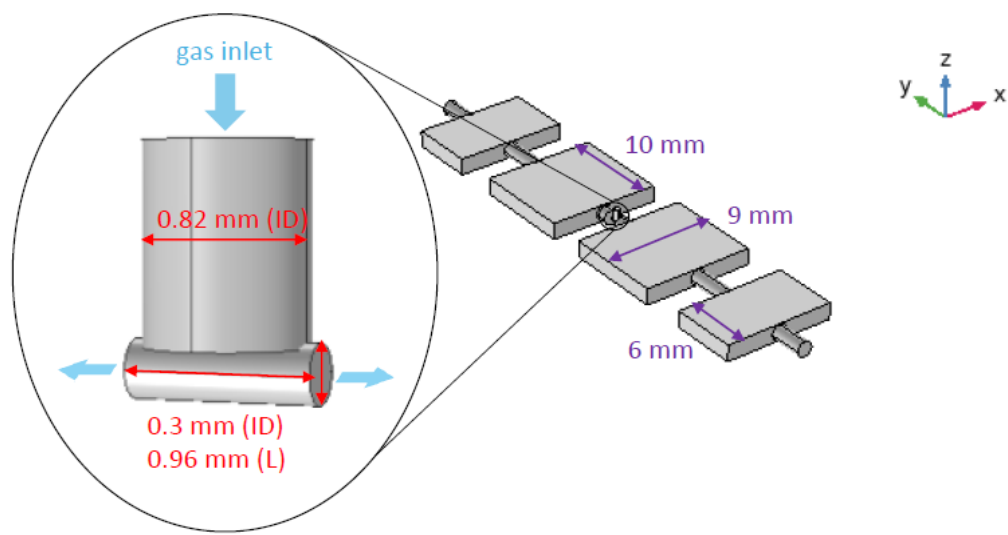

Fig. S1. Schematic of the glass chip dimensions.

To block the residual fundamental beam, metallic filters (Al, Zr, In) were used after generation. The resulting high-order harmonics were then refocused in a 1:1 geometry using a toroidal mirror such to allow the installation of a sample target and further experiments. After the focus, the radiation is then spectrally dispersed with an aberration-corrected concave grating (Hitachi, central groove density 1200 grooves/mm, wavelength range 5–25 nm). The signal was detected using a chevron-stack of microchannel plates (MCP) coupled to a phosphor screen (P43), which was imaged by a CMOS camera placed in air. The overall resolution of the detection system is better than 200 meV. The photon fluxes reported in the main manuscript were measured at the focus position after the toroidal mirror with a SiC-based photodiode optimized for EUV radiation (SCT-EUV20, GaNo Optoelectronics) connected to a lock-in amplifier trigger by the laser itself. Fig. S2 shows an example of uncalibrated HHG spectrum (cutoff region) as recorded at the detector. The figure shows sharp edges (top and bottom of the spectrum), indication that the XUV radiation is partially cut at the toroidal. Quantifying the loss in terms of photon flux though is non-trivial.

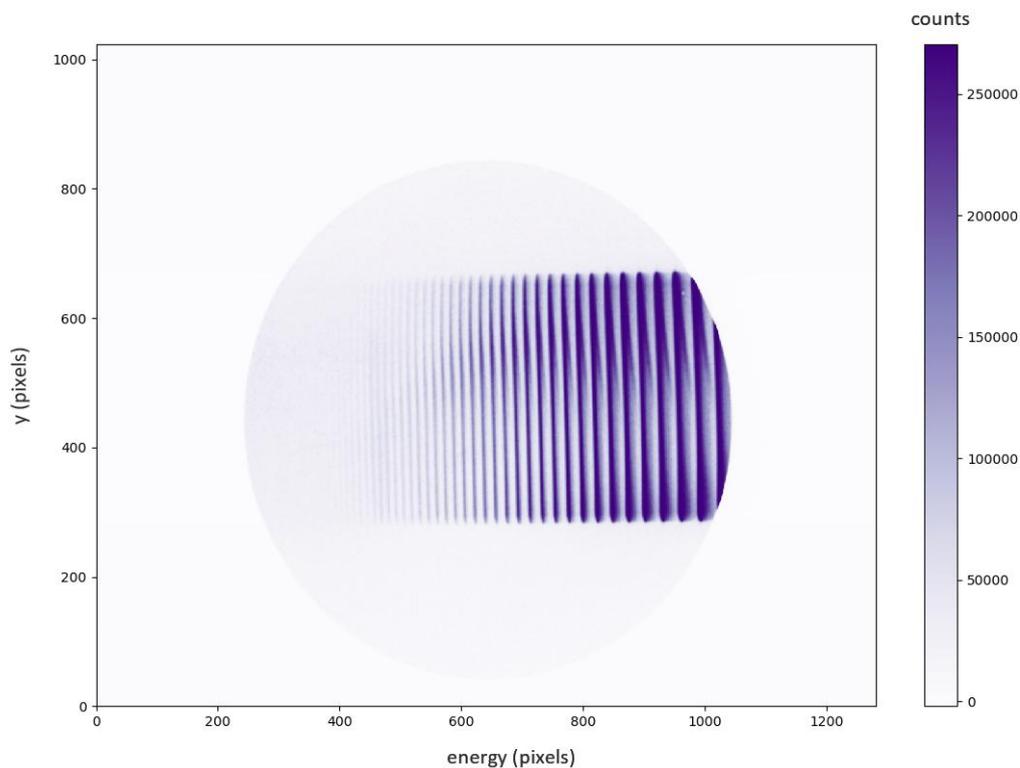

Fig. S2. HHG spatial distribution of the cutoff spectrum in 250 mbar of neon at 800 nm (uncalibrated pixel axes, energy increase going towards the left). Even though the simulations show a near-field confinement of the high energies at outer radii, the measurement shows a rather homogeneous distribution in the far-field, as a result of the energy redistribution at high divergence (the detector is at ca. 3 m from the source point of generation). The sharp horizontal cuts are caused by the limited vertical dimension (15 mm) of the toroidal mirror. Even though near-field simulations show a cutoff generation off-axis, the far-field measurements and simulations don't indicate loss of these energies with propagation.

## S2 - Phase matching: a compendium

For convenience of the readers, this section summarizes on phase matching in HHG. A more complete discussion is provided in Refs. [3,4].

Macroscopic harmonic generation is the result of the coherent sum of the emission contributions from all the atoms in the gas medium. The efficient buildup of the harmonic field relies on the minimization of the phase mismatch between the polarization field at frequency $q\omega_L$, induced by the response of the medium to the fundamental field at a frequency $\omega_L$, and the generated $q$-th harmonic field. The phase mismatch between the two fields is given by:

$$\Delta k = qk_L - k_q = q\left(\frac{n_L \omega_L}{c}\right) - \frac{n_q(q\omega_L)}{c} = \Delta k_{at} + \Delta k_{fe} + \Delta k_{foc} + \Delta k_{dipole}(+\Delta k_{geom}).$$

where the expression for on-axis generation of each term contributing to the phase mismatch is described in the following.

The first term $\Delta k_{at}$ corresponds to the contribution of the neutral atoms and is given by:

$$\Delta k_{at} = (n_L - n_q) \cdot \frac{q\omega_L}{c}$$

with $n_L$ and $n_q$ refractive indices, respectively, at the fundamental frequency $\omega_L$ and at frequency of the $q$-th harmonic above ionization. Employing the definition of polarizability, this term can be rewritten as:

$$\Delta k_{at} = \frac{q\omega_L \rho}{2\varepsilon_0 c} \cdot (1 - \eta_{fe})(\alpha_L - \alpha_q) > 0$$

where $\rho$ is the atomic density in the medium, $\varepsilon_0$ the vacuum permittivity, $c$ is the speed of light in vacuum, $\alpha_L$ and $\alpha_q$ the polarizabilities at the fundamental and $q$-th harmonic frequency, respectively, and $\eta_{fe}$ the ionization fraction in the medium. By approximating the polarizability at the fundamental frequency with the static one and considering that it is greater than $\alpha_q$, this term always contributes positively to the phase mismatch [1].

$\Delta k_{fe}$ corresponds to the dispersion of the free electrons:

$$\Delta k_{fe} = -\frac{q\omega_L \rho}{2\varepsilon_0 c} \cdot \frac{\eta_{fe} e^2}{m_e} \cdot \left(\frac{1}{\omega_L^2} - \frac{1}{q^2 \omega_L^2}\right) < 0$$

with $m_e$ and $e$ electron mass and charge, respectively. This term is negative by definition. The second term in parentheses can be neglected for higher orders.

In the adiabatic approximation (low driver intensity and low ionization level) the terms $\Delta k_{at}$ and $\Delta k_{fe}$ are those mostly contributing to the phase. Perfect phase matching is then obtained when they balance out at the critical ionization level $\eta_{cr}$.

In case of free-space propagation, the Gouy phase term -the shift in phase experienced by the beam while crossing the focus- must be taken into account:

$$\Delta k_{foc} = -\frac{q z_R}{z^2 + z_R^2} < 0$$

with $z_R$ being the Rayleigh range and $z = 0$ corresponding to the position of the focus. This term is negative by defintion. The bigger the distance from the focus, the smaller is this contribution, whilst for short interactions it approaches its maximum value of $-q/z_R$.

In case of guided propagation in capillaries, $\Delta k_{foc} = 0$, but the geometrical phase contribution given by the mode dispersion should be included:

$$\Delta k_{geom} = -q \frac{u_{nm}^2 \lambda_L}{4\pi a^2},$$

with $u_{nm}$ being a coefficient depending on the propagation mode in the capillary (*e.g.* for $J_{11}$, the Bessel function of the first kind and first order, $u_{11} = 2.4$), and $a$ being the core radius of the capillary.

The last term $\Delta k_{dipole}$ corresponds to the dipole phase, which depends on the single atom response and thus on the length of the trajectory that the electron accumulates in the continuum. In the most general form, this term can be expressed as

$$\Delta k_{dipole} = \frac{\partial \Phi_i}{\partial I} \frac{\partial I}{\partial z}$$

To highlight the dependence on the on-axis propagation $z$, it can be rewritten in the following form:

$$\Delta k_{dipole} = -\frac{2z\beta_1(z)}{z^2 + z_R^2}$$

with $\beta_L(z)$ being always a negative coefficient depending on the ionization potential $I_p$, on the harmonic order $q$, and, most importantly, on the intensity and position of the driving field $I_L(z)$ along the propagation axis. The sign of this term depends on the position with respect to focus and it is positive for $z > 0$.

In the nonadiabatic or overdriven regime, the ionization level is high and the primary negative contribution to the phase is due to $\Delta k_{fe}$, whilst the positive contribution of $\Delta k_{at}$ is rather small. The compensation of the negative components (including both the free electron and Gouy phase terms) is mostly due to $\Delta k_{dipole}$. This is why only by placing the medium *after* the focus, thus having $\Delta k_{dipole}$ positive, phase matching in the overdriven regime becomes possible. As highlighted in the main manuscript, plasma, dipole, and spatial phases are dynamically changing along the propagation. Additionally, the strong reshaping prevents an accurate description of the beam as purely gaussian, and its spatial phase cannot be simply defined as the Gouy phase [5].

## S3 - HHG simulations in the overdriven regime

The simulation code employed in this work calculates the response of each atom of the target using the Lewenstein model (strong-field approximation, SFA, and saddle-point method) [6]. The single-atom response is introduced as source term into the Maxwell equations governing the propagation of the fundamental and harmonic beams, which are solved in cylindrical coordinates assuming radial symmetry [7]. For the fundamental beam, the following equation holds:

$$\nabla^2 E_1(r,t,z) - \frac{1}{c^2}\frac{\partial^2 E_1(r,t,z)}{\partial t^2} = \frac{\omega_p^2(r,t,z)}{c^2} E_1(r,t,z)$$

with $\omega_p = \left[\frac{e^2 n_e(r,t,z)}{\epsilon_0 m_e}\right]^{1/2}$ being the plasma frequency depending on the free-electron density $n_e(r,t,z)$ defined by the Lewenstein model and calculated applying the correction to the Ammosov-Delone-Krainov (ADK) model discussed below. The equation is solved in paraxial approximation, taking into account both temporal plasma-induced phase modulation and spatial plasma lensing effects on the fundamental beam. For the harmonic beam, the propagation equation is the following:

$$\nabla^2 E_h(r,t,z) - \frac{1}{c^2}\frac{\partial^2 E_h(r,t,z)}{\partial t^2} = \mu_0 \frac{\partial^2 P_{nl}(r,t,z)}{\partial t^2}$$

where $P_{nl} = [n_0 - n_e(r,t,z)]d_{nl}(r,t,z)$ is the non-linear polarization generated in the gas, $d_{nl}(r,t,z)$ is the non-linear dipole moment as from SFA model, $n_0$ is the neutral atom density and $n_e(r,t,z)$ is the free-electron density, changing along the propagation direction $z$ over time and radial coordinate.

When operating at the high intensities required for the overdriven regime, particular care is required in defining the ionization rate [8]. Under this condition, and particularly for short near-infrared wavelengths, the ionization rate is well described by the above barrier ionization (ABI) model [9], whilst the ADK model [10] typically used to depict the tunnel ionization mechanism breaks down. Ideally, a high-intensity regime would require an exact calculation of the static ionization rates, which, however, is extremely computationally demanding. To overcome this complexity, several approximation or numerical methods have been developed in the years [8,11–13]. We employed the empirically corrected ADK (E-ADK) ionization rate introduced in [13]. Following literature results [11], we further corrected the ionization rate of this model, which is overestimated by a factor two, as follows:

$$w'(t, E_L) = w_{ADK}(t, E_1)\frac{1}{2}exp\left[-\alpha \left(\frac{Z_c^2}{I_p}\right)\left(\frac{E_L(t)}{\epsilon^3}\right)\right]$$

with $w_{ADK}(t, E_1)$ being the ADK rate as in [10], $Z_c = 1$ being the charge seen by the active electron (single ionization), $\alpha = 9$ for Ar and Ne and $\alpha = 7$ for He He obtained by fitting the formula to the ionization rates calculated for a number of atoms and ions within the single-active electron (SAE) approximation [13], $E_L(t)$ being the amplitude of the driving field, $\epsilon = \sqrt{2I_p}$ and $I_p$ being the ionization potential of the gas target. An accurate description of the first ionization rate allows to match our experimental results without the need of including contributions from multiple (sequential and non) ionizations typically considered for high intensities regimes [14–17].

The absorption of the harmonic field was additionally introduced as a factor $e^{-z/2L_{abs}}$, where $L_{abs}$ is the absorption length calculated from the tabulated scattering coefficients for each gas medium [18] and $z$ is the propagation distance. The complex refractive index is expressed by

$$n_x = 1 - \frac{N_{gas}r_e\lambda^2(f_1 - if_2)}{2\pi}$$

with $f_1$ and $f_2$ being the scattering factors, $N_{gas}$ the gas density, $r_e$ the classical electron radius (~2.818 × 10⁻¹⁵ m) and $\lambda$ the wavelength.

Consequently, the absorption coefficient is defined as

$$\sigma_x = \frac{4\pi\, Im\{n_x\}}{\lambda N_{gas}}$$

and the absorption length is defined as

$$L_{abs} = \frac{1}{\sigma_x N_{gas}}$$

### S3.1 Gabor Analysis

The Gabor Analysis [19] was performed as described in [20] to determine the energy content of the harmonic field evolving in time. This allows us to extract information on the dynamic evolution of the generation as a function of time and radial coordinate. The gating window was applied sequentially for each radial coordinate directly to the harmonic field in the temporal domain as follows:

$$E_G(\Omega, t) = \int dt'\, E_h(t') \frac{\exp\left[-(t'-t_0)^2/2\sigma\right]}{\sigma\sqrt{2\pi}} \exp(-i\Omega t')$$

The Gabor field is the result of the Fourier transform of the gated field, using a negative exponential when going from time to spectral domain. The gating function was chosen to be a gaussian function with $\sigma = \frac{1}{3\omega_L}$, sliding of 0.03 fs at each step. The results of this analysis are reported as modulus squared of the Gabor field, providing the HHG intensity. The Gabor maps as function of radial coordinate and time were obtained summing at each radial coordinate the squared modulus of the Gabor field for the interval of energies of interest (between 200 and 220 eV in **Fig. 4b** and **5**).

The analysis was performed also using the Hann function as alternative gating function to the Gaussian one, obtaining similar results and showing the robustness of the method.

## S3.2 – Extended simulation results: generation in neon, 800-nm driver

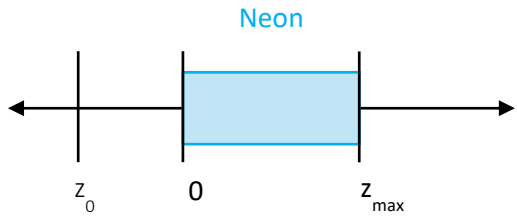

| Parameters | |
|---|---|
| Distance focus entrance to the medium $z_0$ | -900 µm |
| Medium length $z_{max}$ | 900 µm |
| Spot size at focus $w_0$ | 20 µm |
| Rayleigh length $z_R$ | 1.57 mm |
| Duration $\Delta t_{FWHM}$ | 30 fs |
| Field amplitude $A_0$ | 0.5338 a.u. ($10^{16}$ W/cm²) |
| Number gas density | 9·$10^{-7}$ a.u. (250 mbar) |
| Driving frequency $\omega_0$ | 0.0570 a.u. (800 nm) |

The integrated spectrum (Fig. S3a) reproduces well the expected cutoff at 200 eV. The discrepancy in the intensity with respect to the experiment is due to a known underestimation of the SFA model in determining the harmonic strength up to even a factor ten [8]. What is crucial in assessing the correct description of the overdriven regime is the reshaping of the driving field, which is fully captured by the current simulations. Fig. S3c reports the spectral reshaping of the driving field at the end of the medium, with a pronounced shift towards high frequency (shorter wavelengths) and a peaked intensity at outer radii. For comparison, Fig. S3b reports the initial spectrum of the driving field, *i.e.* a gaussian pulse of 30 fs centered at 800 nm. Fig. S3d highlights the on-axis temporal reshaping (black curve), characterized by a shrinking of the intensity and a temporal chirp of the leading edge of the pulse with respect to the initial driving field (red curve). Based on the Gabor analysis reported in Fig. 4 of the main text, the temporally reshaped fraction of the pulse, *i.e.*, its leading edge, drives the generation at high energies, suggesting that the phase matching is enhanced by the phase distortions.

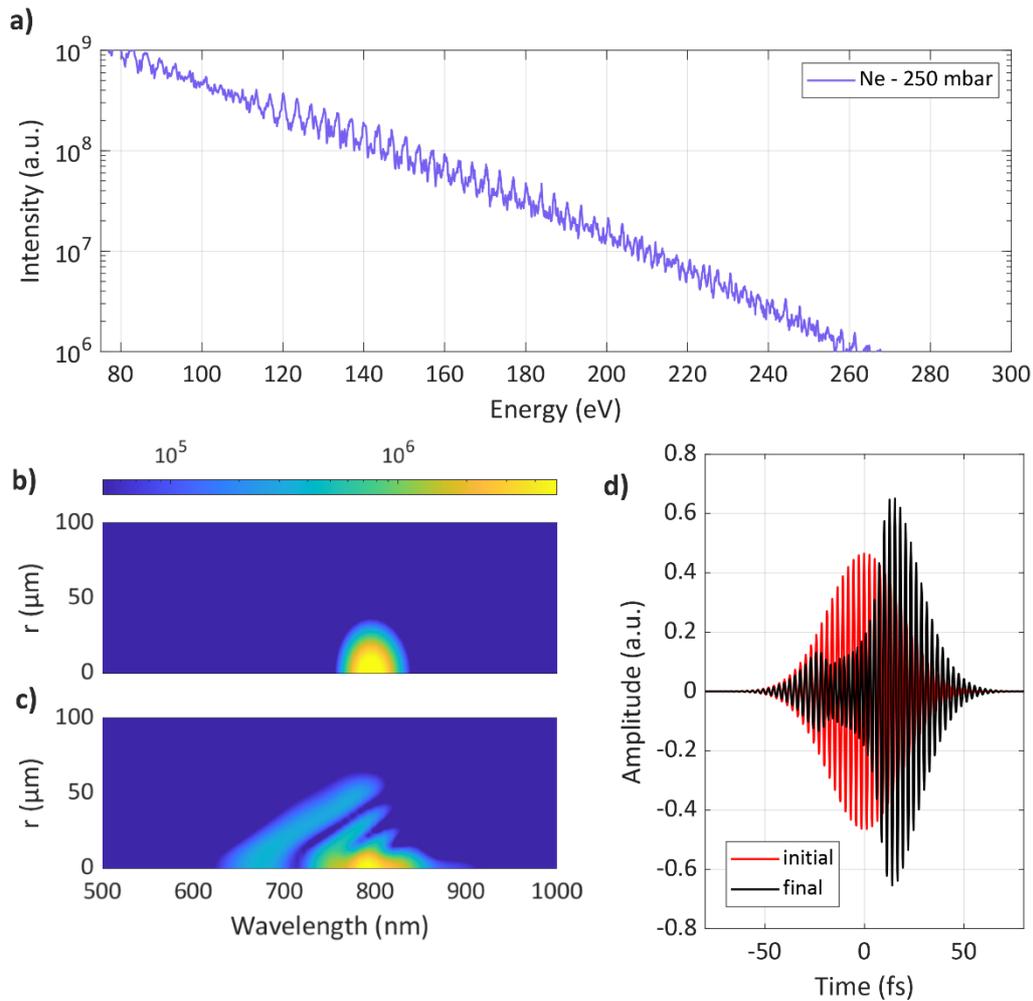

Fig. S3. Simulated three-dimensional space generation and propagation of the 800 nm driving field and HHG field in Ne. The medium is located 900 μm mm after the focus of the beam, corresponding to approximately one Rayleigh length. (a) Integrated harmonic spectrum over the radial coordinate. (b-c) Spectral intensity (squared module) of the driving field respectively at the beginning and at the end of the propagation medium as function of the radial coordinate r and the driving wavelength. (d) On-axis (r=0) temporal amplitude of the driving field at the beginning (red) and at the end (black) of the propagation medium. All the quantities are expressed in atomic units (a.u.).

Figure S4a shows the calculated harmonic spectrum generated by the 800-nm driver in neon and integrated over the radial coordinate as a function of an extended propagation distance of 2700 μm (as Fig. 3a in the main manuscript). Figure S4b shows the harmonic yield as function of the propagation distance in the gas medium for three energies of the plateau region, 80 eV (light blue), 100 eV (purple), and 120 eV (orange). Differently from the cutoff energies reported in Figure 3b of the main manuscript, the harmonic yield at these low photon energies increases up to 2 mm and suddenly drops of almost an order of magnitude due to reabsorption.

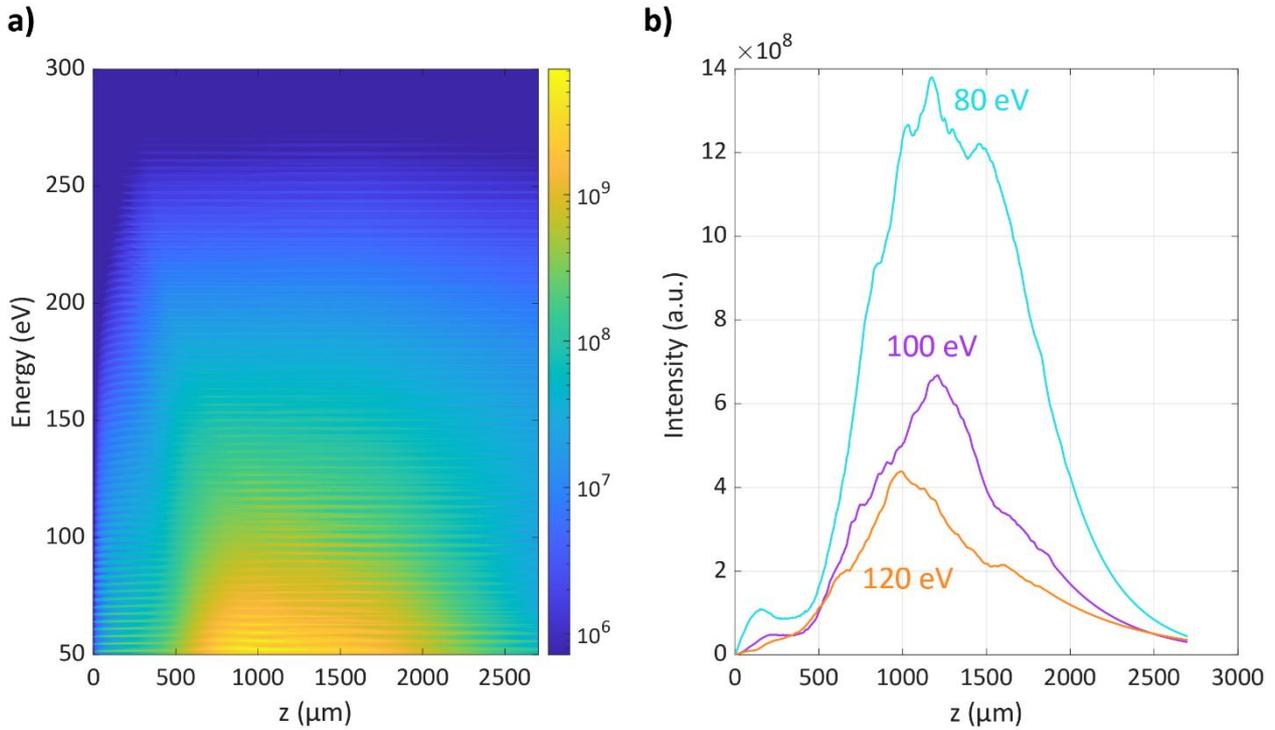

Fig. S4. Simulated three-dimensional space generation and propagation of the HHG spectrum in Ne with 800 nm driving field over 2.7 mm. (a) HHG field spectral intensity integrated over the radial coordinate as function of the propagation (z-axis) in the gas medium. (b) Intensity in atomic units for three energies (80 eV, 100 eV, 120 eV) in the plateau region of the spectrum as function of the propagation distance in the gas medium. For these low energies, saturation is reached between 1 and 1.5 mm followed by strong reabsorption for the remaining part of the propagation, with an intensity decrease of more than an order of magnitude. Contrary to the propagation of high energies shown in the main manuscript, here there is no region for which the yield is preserved, and re-absorption becomes dominant right after saturation is reached.

The driving field spatiotemporal profile for different positions along the propagation in the gas medium is shown in Figure S5. The intensity peak progressively shifts towards later time cycles. This effect allows for efficient phase matching of higher harmonics over multiple cycles, extending the cutoff emission. At the same time, the high on-axis intensity causes the full ionization of the gas medium, preventing efficient phase matching. Based on these plots and Figure 5 in the main manuscript, at the end of the medium the generation occurs over the full leading edge of the pulse, *i.e.*, for all the cycles corresponding to negative times.

The corresponding ionization degree (1=medium fully ionized, 0=not ionized) of the medium is reported in Figure S6. For longer propagation, the region of full ionization shifts towards later time cycles, *i.e.*, the trailing edge of the pulse, and shrinks to lower radii, allowing more driving field cycles to contribute to the efficient generation of harmonics, progressively at smaller radii (see Fig. 5 in the main manuscript).

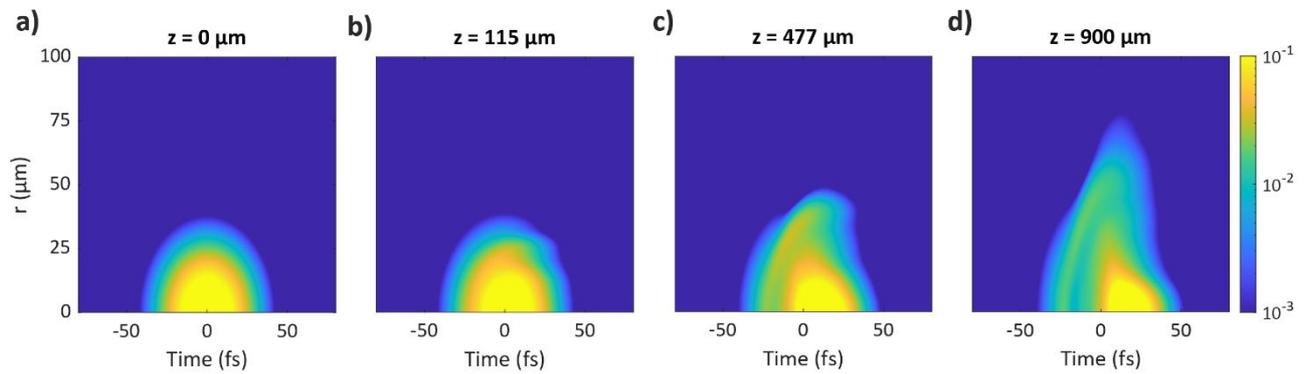

Fig. S5. Simulated driving intensity in atomic units of intensity (log-scale) as a function of time and radial coordinate r for four different points along the propagation direction z, where z = 0 µm corresponds to the beginning of the gas medium. The beam has a spot size of 20 µm at z = -900 µm. The spatial reshaping of the field starts in the first hundred microns of propagations, while the temporal reshaping (intensity shift towards the trailing edge) occurs in the second half of the propagation medium. The combination of both effects results in the final intensity profile of panel d, characterized by high intensity at late cycles and by an extended spatial distribution beyond 50 µm.

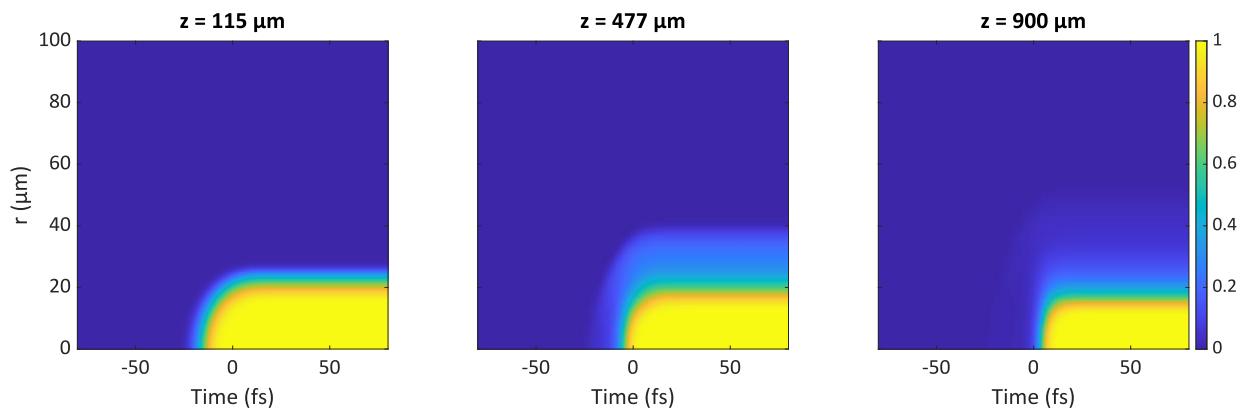

Fig. S6. Simulated ionization degree of the gas medium for three different positions along the propagation axis. The region in which the medium is fully ionized progressively reduces to lower coordinates and shifts to later times.

Figure S7 reports the HHG intensity as function time and energy in *argon* for a radial coordinate of 23 μm corresponding to the maximum cutoff emission, as obtained through Gabor analysis. The generation conditions are the same as in neon (see table at the beginning of the section), with optimal pressure set to 200 mbar and medium position to -1.25 mm, corresponding to the experimental conditions. For time cycles between -10 to -8, HHG of the cutoff energies above 100 eV is dominated by short trajectories. For later cycles, long trajectories become instead dominant. With a similar dynamic to neon, the generation of the cutoff extends for multiple cycles over the leading edge of the driving field.

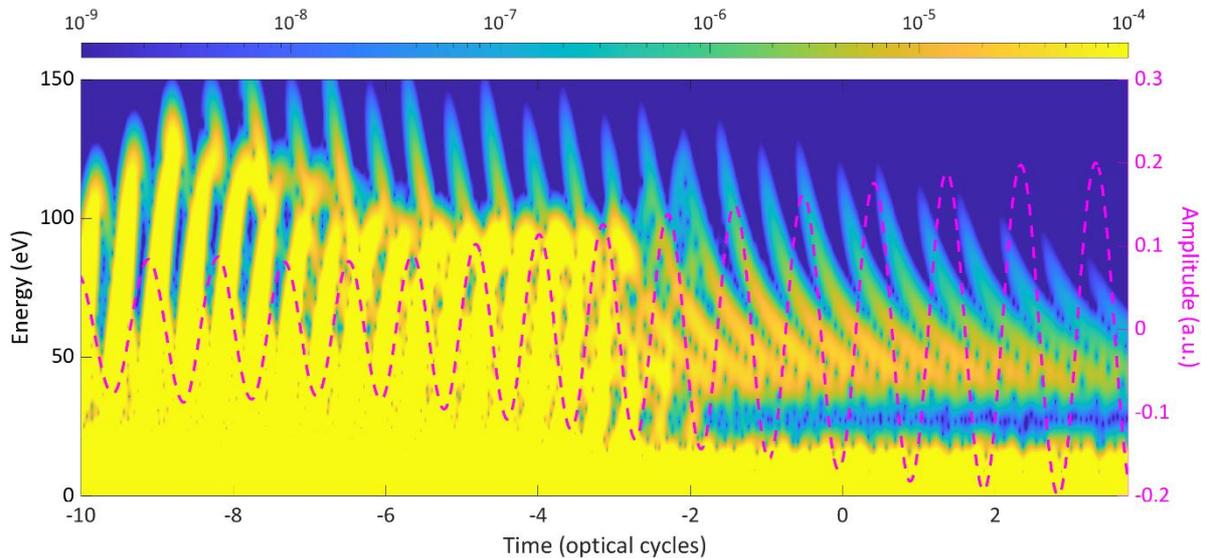

Fig. S7. Gabor analysis for the HHG spectrum (a.u.) in argon with 800 nm driving wavelength as a function of the photon energy and optical cycle for a radial coordinate of 23 μm, corresponding to conditions of maximum extension of the photon energy cutoff. The driving field amplitude is reported in pink in atomic units.